\documentclass[pra,twocolumn,showpacs,preprintnumbers,amsmath,amssymb]{revtex4}
\usepackage{mathrsfs}
\usepackage{bbm}
\usepackage{amsfonts}
\usepackage{tipa}

\usepackage{epsfig,graphicx}
\usepackage{amstext}
\usepackage{amsmath}
\usepackage{graphicx}
\usepackage{times}

\begin{document}


\title{Geometric quantum discord and non-Markovianity of structured reservoirs}

\author{Ming-Liang Hu}
\email{mingliang0301@163.com}
\author{Han-Li Lian}
\affiliation{School of Science, Xi'an University of Posts and Telecommunications,
             Xi'an 710061, China}

\begin{abstract}
The reservoir memory effects can lead to information backflow and
recurrence of the previously lost quantum correlations. We establish
connections between the direction of information flow and variation
of the geometric quantum discords (GQDs) measured respectively by
the trace distance, the Hellinger distance, and the Bures distance
for two qubits subjecting to the bosonic structured reservoirs, and
unveil their dependence on a factor whose derivative signifies the
(non-)Markovianity of the dynamics. By considering the reservoirs
with Lorentzian and Ohmic-like spectra, we further demonstrated that
the non-Markovianity induced by the backflow of information from the
reservoirs to the system enhances the GQDs in most of the parameter
regions. This highlights the potential of non-Markovianity as a
resource for protecting the GQDs.

\end{abstract}

\pacs{03.65.Ud, 03.65.Ta, 03.67.Mn
      \\Key Words: Geometric quantum discord; non-Markovianity; Structured reservoir
     }

\maketitle

\section{Introduction}\label{sec:1}
Quantum correlations occupy an important position in understanding
fundamental characteristics of a quantum system. For this reason,
they remain the research focus of people from the early days of
quantum mechanics to now. Today, when we mention to quantum
correlations, we know that in addition to entanglement
\cite{rmp-en}, the concept of quantum discord constitutes another
representative class of quantum correlation measure \cite{qd-von}.
The related studies on this subject are mainly carried out around
its quantification, its particular behaviors, and the control of it
in various quantum systems \cite{rmp-qd}. Particularly, there has
been an increasing interest of people on quantifying quantum discord
from different perspectives, and to date there are a number of
discord-like correlation measures being proposed \cite{qd-von,
mid,qd-hil,gqd-tra,gqd-hel,gqd-lqu,gqd-bur}. On the other hand, the
behaviors of quantum discord in the spin chain \cite{spin}, the
two-level atoms \cite{atom}, and the NMR system \cite{NMR} have also
been studied extensively.

From an applicative point of view, quantum discord is an invaluable
resource for implementing many quantum tasks \cite{DQC1,rsp,inf,
qsm,qsb,metro}. But it is very fragile, and the unavoidable
interaction of a realistic system with its environment leads to
irretrievable deterioration of the correlations in most cases
\cite{robu1,robu2,robu3,robu4,robu5,robu6}. This makes understanding
of the connection between the environmental effects and evolution of
quantum discord a vital problem. In fact, many studies have already
been performed in this respect, and there were evidence indicating
that sometimes the non-Markovian character of an environment may
serve as a resource for protecting quantum discord from being
destroyed completely \cite{Markov1,Markov2, Markov3,Markov4}. It has
also been observed that with elaborately chosen spectrum of the
reseroir, the quantum discord can be frozen for an interval of time
\cite{robu4,froz} or be frozen permanently \cite{qd-Markov1}.

Although it is evident that sometimes the non-Markovianity can be
used to enhance quantum discord of a system to some extent, we must
to say that this is not always the case \cite{qd-Markov1,qd-Markov2,
qd-Markov3}. Searching a general connection between non-Markovian
character of an environment and the variation tendency of quantum
discords is still an open subject in the quest for reliable ways to
protect them. Toward that end, in this paper we establish an
explicit dependence of the geometric quantum discords (GQDs)
\cite{gqd-tra, gqd-hel, gqd-lqu,gqd-bur} of a two-qubit system on
non-Markovianity of the zero-temperature bosonic structured
reservoirs, and unveil the connections between the direction of
information flow and enhancement of the GQDs for different initial
states. Actually, with the rapid developments of the reservoir
engineering technique \cite{RET1,RET2, RET3}, nowadays it is
feasible to adjust experimentally frequency distribution of a
reservoir to the desired regime such that the decay time for the
quantum discord can be prolonged, provided that we know the explicit
dependence of it on spectral density distribution of the reservoir.

The structure of this paper is arranged as follows. In Sec.
\ref{sec:2} we recall briefly measures of the GQDs, while in Sec.
\ref{sec:3} the model for the system-reservoir coupling is
presented. Sec. \ref{sec:4} is devoted to a derivation of the GQDs
and their dependence on a reservoir-determined factor. Then in Sec.
\ref{sec:5}, we illustrate via two explicit examples our main
findings. Finally, Sec. \ref{sec:6} is devoted to a summary.

\section{Measures of the GQD}\label{sec:2}
There are many discord measures being proposed until now. We recall
here three measures of the GQD. They are defined, respectively, by
the trace distance, the Hellinger distance, and the Bures distance
\cite{gqd-tra,gqd-hel,gqd-lqu, gqd-bur}. For simplicity, we will
call them the trace distance discord (TDD), the Hellinger distance
discord (HDD), and the Bures distance discord (BDD).

To begin with, we list some notations we used. We denote by $\rho$
the density operator of a bipartite system $AB$, and $\Omega_0 =
\sum_k p_k \Pi_k^A\otimes \rho_k^B$ the set of zero-discord states
\cite{qd-hil}, with $\Pi_k^A$ the orthogonal projector in the
Hilbert space $\mathcal {H}_A$, and $\rho_k^B$ an arbitrary density
operator in $\mathcal {H}_B$, $0\leq p_k \leq1$ and $\sum_k p_k=1$.
Moreover, $|| X ||_p =[\mbox{Tr} (X^\dag X) ^{p/2}]^{1/p}$ is the
Schatten $p$-norm, which reduces to the trace norm for $p=1$, and
the Hilbert-Schmidt norm for $p=2$.

The first GQD measure we considered is the well-accepted TDD. Its
definition is as follows \cite{gqd-tra}
\begin{equation}\label{eq2-1}
 D_{\rm T}(\rho)=\min_{\chi\in\Omega_0}\parallel\rho-\chi\parallel_1,
\end{equation}
and for the two-qubit states $\rho^X$ with the {\it X}-shaped matrix
form (i.e., $\rho^X$ contains nonzero elements only along the main
diagonal and anti-diagonal), the TDD can be obtained analytically
\cite{trace-ana}. Particularly, for a restricted subset of $\rho^X$
with elements $\rho_{14,41}^X =0$, we have \cite{trace-ana}
\begin{eqnarray}\label{eq2-2}
 D_{\rm T}(\rho^X)=2|\rho_{23}^X|.
\end{eqnarray}

The second GQD measure is the HDD, which is a modified version of
the earliest proposed GQD \cite{qd-hil}. It reads \cite{gqd-hel}
\begin{equation}\label{eq2-3}
 D_{\rm L}(\rho)=2\min_{\Pi^A}\parallel\sqrt{\rho}-\Pi^A(\sqrt{\rho})\parallel_2^2,
\end{equation}
where the minimum is taken over $\Pi^A = \{\Pi_k^A\}$, with
\begin{equation}\label{eq2-4}
 \Pi^A(\sqrt{\rho})= \sum_k (\Pi_k^A\otimes I_B)\sqrt{\rho}(\Pi_k^A\otimes I_B),
\end{equation}
and $I_B$ is the identity operator in $\mathcal {H}_B$.
Particularly, if we are restricted to the $(2\times n)$-dimensional
$\rho$, Eq. \eqref{eq2-3} yields \cite{gqd-lqu}
\begin{eqnarray}\label{eq2-5}
 D_{\rm L}(\rho)=1-\lambda_{\max}\{W_{AB}\},
\end{eqnarray}
where $\lambda_{\max}\{W_{AB}\}$ is the maximum eigenvalue of the
matrix $W_{AB}$ whose elements are given by
\begin{eqnarray}\label{eq2-6}
 (W_{AB})_{ij}={\rm Tr}\{\sqrt{\rho}(\sigma_i^A\otimes I_B)
               \sqrt{\rho}(\sigma_j^B\otimes I_B)\},
\end{eqnarray}
with $\sigma_{1,2,3}^S$ ($S=A,B$) the three Pauli operators.

Finally, we recall the GQD measure of BDD,
which is defined as \cite{gqd-bur}
\begin{equation}\label{eq2-7}
 D_{\rm B}(\rho)=\sqrt{(2+\sqrt{2})[1-\max_{\chi\in\Omega_0}\sqrt{F(\rho,\chi)}]},
\end{equation}
with $F(\rho,\chi)= [{\rm Tr}(\sqrt{\rho}\chi\sqrt{\rho})^{1/2}]^2$,
and for the special case of $(2\times n)$-dimensional state $\rho$,
$F_{\max} (\rho,\chi)=\max_{\chi\in\Omega_0} F(\rho,\chi)$
simplifies to \cite{bures-ana}
\begin{eqnarray}\label{eq2-8}
 F_{\max}(\rho,\chi)=\frac{1}{2}\max_{||\vec{u}=1||}
 \left(1-{\rm Tr}\Lambda+2\sum_{k=1}^{n_B}\lambda_k(\Lambda)\right),
\end{eqnarray}
with $\lambda_k(\Lambda)$ being the eigenvalues of
$\Lambda=\sqrt{\rho} (\vec{u}\cdot\vec{\sigma}^A\otimes I_B)
\sqrt{\rho}$ arranged in non-increasing order, $n_B=\dim \mathcal
{H}_B$, and $\vec{u}$ a unit vector in $\mathbb{R}^3$.

\section{The model}\label{sec:3}
We consider in this paper two noninteracting qubits denoted by $S=A$
and $B$. Each of them coupled locally to their independent
zero-temperature bosonic reservoir. The Hamiltonian of the ``qubit
plus reservoir'' subsystem reads \cite{master}
\begin{equation}\label{eq3-1}
 {\hat H}=\omega_0\sigma_{+}\sigma_{-}+\sum_{k}\omega_k b_k^{\dag} b_k
          +\sum_{k}(g_k b_k\sigma_{+}+{\rm H.c.}),
\end{equation}
with $\omega_0$ being the transition frequency of the qubit, and
$\sigma_\pm$ the raising and lowering operators. $b_k$ and
$b_k^{\dag}$ are the annihilation and creation operators for the
field mode $k$ with frequency $\omega_k$ and the system-reservoir
coupling constant $g_k$.

When the initial state of each qubit with its reservoir is in a
product form, the evolution of the reduced density matrix for qubit
$S$ is known to be described by \cite{master}
\begin{eqnarray}\label{eq3-2}
 \dot{\rho}^{S}(t)&=&-i\frac{\Omega(t)}{2}[\sigma_{+}\sigma_{-},\rho^{S}(t)]+
                     \frac{\Gamma(t)}{2}[2\sigma_{-}\rho^{S}(t)\sigma_{+}\nonumber\\
                  &&-\{\sigma_{+}\sigma_{-},\rho^{S}(t)\}],
\end{eqnarray}
where the time-dependent factors $\Gamma(t)$ and $\Omega(t)$ are as
follows
\begin{eqnarray}\label{eq3-3}
 \Gamma(t)=-2{\rm Re}\left[\frac{\dot{p}(t)}{p(t)}\right],~
 \Omega(t)=-2{\rm Im}\left[\frac{\dot{p}(t)}{p(t)}\right],
\end{eqnarray}
with ${\rm Re}[x]$ and ${\rm Im}[x]$ representing, respectively, the
real and imaginary parts of $x$, while $p(t)$ obeys the
integro-differential equation \cite{Bellomo}
\begin{equation}\label{eq3-4}
 \dot{p}(t)+i\omega_0 p(t)+\int_0^t p(t_1)f(t-t_1)dt_1=0,
\end{equation}
where the correlation function $f(t-t_1)$ is related to the
spectral density $J(\omega)$ of the reservoir via
\begin{equation}\label{eq3-5}
f(t-t_1)=\int d\omega J(\omega)e^{-i\omega(t-t_1)}.
\end{equation}

From Eq. \eqref{eq3-2} one can show that the reduced density matrix
$\rho^S(t)$ for qubit $S$ takes the form \cite{Bellomo}
\begin{equation}\label{eq3-6}
 \rho^S(t)=\left(\begin{array}{cc}
           \rho^S_{11}(0)|p(t)|^2  & \rho^S_{10}(0)p(t) \\
           \rho^S_{01}(0)p^*(t)    & 1-\rho^S_{11}(0)|p(t)|^2
    \end{array}\right),
\end{equation}
with $\rho^S_{ij}(0)=\langle i|\rho^S(0)|j\rangle$, and
$\{|1\rangle,|0\rangle\}$ the standard basis.

If we further define $q(t)=|p(t)|^2$, then the decay rate
$\Gamma(t)$ in Eq. \eqref{eq3-3} turns out to be
\begin{equation}\label{eq3-7}
 \Gamma(t)=-\frac{\dot{q}(t)}{q(t)},
\end{equation}
whose sign is determined solely by the slope of $q(t)$, namely, by
$\dot{q}(t)=\partial q(t)/\partial t$.

The sign of $\Gamma(t)$ is also intimately related
to the direction of information flow between the system and the
reservoir \cite{Mark1,Mark2,Mark3,Mark4}. If $\Gamma(t)$ is always
positive, i.e., $\Gamma(t)> 0$ in the whole time region, the
evolution process is said to be Markovian, and the information flows
from the system into the reservoir. On the other hand, it is
non-Markovian if $\Gamma(t)$ takes on negative values within certain
time intervals, and now there are temporary information backflow
from the reservoir to the system. By the way, as the energy
$\varepsilon(t) = {\rm Tr}[\rho^S (t) H_S]$ ($H_S=\omega_0
\sigma_{+} \sigma_{-}$) for qubit $S$ was given by $\varepsilon(t)
= \omega_0\rho^S_{11}(0) q(t)$, the information backflow is also
accompanied by the energy backflow.

In this paper, we will show that the direction of information flow
between the system and the reservoir is also intimately related to
the variation tendency of the GQDs. Particularly, the
non-Markovianity of the dynamics can be detected efficiently by
tracking the evolution of the GQDs.

\section{Connection between enhancement of GQDs and direction of information flow} \label{sec:4}
We consider two qubits being prepared initially in pure state of the
following form
\begin{equation}\label{eq4-1}
 |\Phi\rangle=\alpha|10\rangle+\sqrt{1-\alpha^2}|01\rangle,
\end{equation}
for which the two-qubit density matrix $\rho(t)$ is given by
\begin{equation}\label{eq4-2}
 \rho(t)=\left(\begin{array}{cccc}
           0  &  0               &  0               & 0 \\
           0  & \alpha^2 q(t)    & \alpha\beta q(t) & 0 \\
           0  & \alpha\beta q(t) & \beta^2 q(t)     & 0 \\
           0  & 0                & 0                & 1-q(t)
    \end{array}\right),
\end{equation}
where $\beta=\sqrt{1-\alpha^2}$. Clearly, $\rho(t)$ maintains the
$X$ form, and the GQDs are determined by the time-dependent factor
$q(t)$. Moreover, one can show that the three GQDs are independent
of the sign of $\alpha$, so we consider in the following only the
initial state $|\Phi\rangle$ with $\alpha\geqslant 0$.

\subsection{The case of TDD}
For $\rho(t)$ in Eq. \eqref{eq4-2}, the TDD can be obtained as
\begin{equation}\label{eq4-3}
 D_{\rm T}(\rho)=2\alpha \beta q,
\end{equation}
which behaves as a monotonic increasing function of $q$, except the
trivial cases of $\alpha^2 = 0$ and 1.

This result indicates that if one can engineer spectral distribution
of the structured reservoir such that the time-dependent factor
$q(t)$ is increased in certain time intervals, the TDD can be
enhanced, and its maximum is achieved when $q(t)$ reaches its
maximum. Moreover, from Eq. \eqref{eq3-7} one can see that the
increase of $q(t)$ with time corresponds to the negative
$\Gamma(t)$. Therefore, for the present case the negative decay rate
$\Gamma(t)$, or equivalently, the backflow of information from the
reservoirs to the system, is always favorable for enhancing the TDD.

\subsection{The case of HDD}
We now turn to consider the HDD. For $\rho(t)$ in Eq. \eqref{eq4-2},
the eigenvalues of $W_{AB}$ can be derived as
\begin{equation}\label{eq4-4}
  \lambda_{1,2}=2\alpha^2\sqrt{q(1-q)},~~
  \lambda_3=1-4\alpha^2\beta^2 q,
\end{equation}
the relative magnitudes of which depend on the parameters involved,
and thus the analytical solution of $D_{\rm L}(\rho)$ is somewhat
complex. We discuss it via the following two cases.

\begin{figure}
\centering
\resizebox{0.43\textwidth}{!}{%
\includegraphics{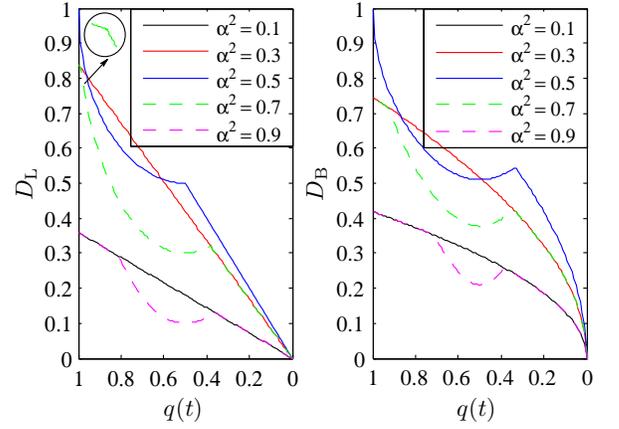}}
\caption{(Color online) The HDD $D_{\rm L}$ and BDD $D_{\rm B}$
         versus $q(t)$ with different values of the initial
         state parameter $\alpha^2$.} \label{fig:1}
\end{figure}

First, for $\alpha^2 \leqslant 1/3$, we have $\lambda_{\max} (W_{AB})=
\lambda_3$, thus
\begin{equation}\label{eq4-5}
 D_{\rm L}(\rho)=4\alpha^2\beta^2 q,
\end{equation}
which increases with the increase of $q$ except the trivial case of
$\alpha^2 =0$. See, for example, the plots of $D_{\rm L}(\rho)$
versus $q$ for $\alpha^2=0.1$ and 0.3 showed in the left panel of
Fig. \ref{fig:1}.

Second, for $\alpha^2 > 1/3$, we have
\begin{equation}\label{eq4-6}
  D_{\rm L}(\rho)=\left\{
  \begin{aligned}
   & 1-2\alpha^2\sqrt{q(1-q)} & \text{if}~ q \in    [q_{c1},q_{c2}],\\
   & 4\alpha^2\beta^2 q   & \text{if}~     q \notin [q_{c1},q_{c2}],
  \end{aligned} \right.
\end{equation}
where the parameters $q_{c1}$ and $q_{c2}$ are given by
\begin{equation}\label{eq4-7}
 q_{c1,c2}=\frac{(2-\alpha^2) \mp \beta\sqrt{3\alpha^2-1}}
           {2\alpha^2(1+4\beta^4)},
\end{equation}
and $q_{c1}$ takes the $``-"$ sign, $q_{c2}$ takes the $``+"$ sign.
One can check that for $\alpha^2 \in (1/3, 0.5)$, both $q_{c1}$ and
$q_{c2}$ are larger than 0.5, while for $\alpha^2 \in (0.5, 1)$, we
have $q_{c1}\leqslant 0.5$ and $q_{c2}\geqslant 0.5$. Moreover, for
$\alpha^2=0.5$, we have $q_{c1}=0.5$ and $q_{c2}=1$.

From Eq. \eqref{eq4-6} one see that $D_{\rm L}(\rho)$ is a monotonic
increasing function of $q$ when $q\in [q_{c1},q_{c2}]$ with $q_{c1}
\geqslant 0.5$, and when $q \notin [q_{c1}, q_{c2}]$, while it is a
monotonic decreasing function of $q$ otherwise. By combing these
with Eq. \eqref{eq4-7}, we summarize the $q$ dependence of $D_{\rm
L}(\rho)$ as follows:

(i) If $\alpha^2 \in (1/3, 0.5]$, $D_{\rm L}(\rho)$ always behaves
as a monotonic increasing function of $q$, as exemplified by the
blue curve for $\alpha^2=0.5$ in the left panel of Fig. \ref{fig:1}.

(ii) If $\alpha^2\in(0.5,1)$, $D_{\rm L}(\rho)$ is a monotonic
increasing function of $q$ in the regions $q\leqslant q_{c1}$ and $q
\geqslant 0.5$, and a monotonic decreasing function of $q$ in the
region $q\in (q_{c1}, 0.5)$. See, the exemplified plot for $\alpha^2
= 0.7$ and 0.9 showed in Fig. \ref{fig:1}.

From the above discussion we see that for the initial states
$|\Phi\rangle$ with $\alpha^2\leqslant 0.5$, the HDD can always be
enhanced by the backflow of information from the reservoir to the
system. For $\alpha^2 > 0.5$, however, the HDD is enhanced by the
information backflow only when $q\leqslant q_{c1}$ and $q \geqslant
0.5$, while it is enhanced with the information loss when $q\in
(q_{c1}, 0.5)$.

\subsection{The case of BDD}
When considering the BDD for the initial state $|\Phi\rangle$, by
writing the unit vector $\vec{u}=(\sin\theta\cos\phi, \sin\theta
\sin\phi,\cos\theta)$, one can derive the eigenvalues of $\Lambda$
analytically as
\begin{equation}\label{eq4-8}
  \begin{aligned}
   & \epsilon_{1,2}=0,\\
   & \epsilon_{3,4}=\frac{1}{2}[\chi\cos\theta\pm
                    \sqrt{\xi\cos^2\theta+4\alpha^2 q(1-q)}],
  \end{aligned}
\end{equation}
where $\chi=2\alpha^2 q-1$, and $\xi=4(1-\alpha^2\beta^2)q^2
-4q+1$.

Due to the parameters $\alpha^2$, $\cos\theta$, and $q$ involved,
the $\epsilon_i$ cannot be ordered by magnitude in general. But as
$\epsilon_3 \geqslant \epsilon_4$, solutions of $F_{\max}$ in
$D_{\rm B}(\rho)$ can be obtained by separating $\alpha^2$ into the
following three different regions.

First, for $\alpha^2\leqslant 1/3$, we have
\begin{equation}\label{eq4-9}
 F_{\max}=1- \alpha^2 q,
\end{equation}
by combining of which with Eq. \eqref{eq2-7}, one can note that
except the trivial case $\alpha^2=0$, $D_{\rm B}(\rho)$ always
increases with the increase of $q$. See, e.g., the exemplified plots
for $\alpha^2=0.1$ and 0.3 displayed in the right panel of Fig.
\ref{fig:1}.

Second, for $\alpha^2\in(1/3,0.5]$, we have
\begin{equation}\label{eq4-10}
  F_{\max}=\left\{
  \begin{aligned}
   & 1- \alpha^2 q,   & \text{if}~ q \notin [q_{c3},q_{c4}],\\
   & \frac{1}{2}+\sqrt{\alpha^2 q(1-q)} & \text{if}~ q \in [q_{c3},q_{c4}],
  \end{aligned} \right.
\end{equation}
where the parameters
\begin{equation}\label{eq4-11}
  q_{c3,c4}=\frac{2\alpha\mp\sqrt{3\alpha^2-1}}{2\alpha(1+\alpha^2)},
\end{equation}
and $q_{c3}$ decreases from 0.75 to $1/3$, while $q_{c4}$ increases
from 0.75 to $1$. Then, in the regions of $q < q_{c3}$ and $q >
q_{c4}$, $F_{\max}$ is decreased by increasing $q$. In the region
of $q \in [q_{c3}, q_{c4}]$, however, the situation is somewhat
complicated: if $q_{c3}\geqslant 0.5$, which corresponds to $\alpha^2 \in
(1/3,0.382]$, $F_{\max}$ is decreased by increasing $q$; if $q_{c3}<
0.5$, which corresponds to $\alpha^2 \in (0.382,0.5]$, $F_{\max}$ is
increased (decreased) by increasing $q$ when $q\in[q_{c3},0.5]$
($q\in(0.5, q_{c4}]$). Thus, the $q$ dependence of $D_{\rm B}(\rho)$
are as follows:

(i) If $\alpha^2 \in (1/3,0.382]$, $D_{\rm B}(\rho)$ always
increases with the increase of $q$.

(ii) If $\alpha^2 \in (0.382,0.5]$, $D_{\rm B}(\rho)$ increases
(decreases) with the increase of $q$ when $q<q_{c3}$ and $q> 0.5$
($q\in[q_{c3},0.5]$). See the blue curve for $\alpha^2=0.5$ in the
right panel of Fig. \ref{fig:1}

Finally, for $\alpha^2 > 0.5$, we have
\begin{equation}\label{eq4-12}
  F_{\max}=\left\{
  \begin{aligned}
   &\frac{1}{2}+\sqrt{\alpha^2q(1-q)}, & \text{if}~ q \in [q_{c5},q_{c6}],\\
   &\frac{1}{2}[1+ \sqrt{\gamma+4\alpha^2 q(1-q)}] & \text{if}~ q \notin [q_{c5},q_{c6}],\\
  \end{aligned} \right.
\end{equation}
with the parameters $q_{c5}$ and $q_{c6}$ being given by
\begin{equation}\label{eq4-13}
  q_{c5,c6}=\frac{1\mp\alpha\beta}{2(1-\alpha^2\beta^2)},
\end{equation}
and $q_{c5}$ increases from $1/3$ to $0.5$, while $q_{c6}$ decreases
from $1$ to $0.5$. Then, by combining this with Eq. \eqref{eq2-7}
one can obtain that $D_{\rm B}(\rho)$ is a monotonic increasing
function of $q$ in the regions $q < q_{c5}$ and $q > 0.5$, and a
monotonic decreasing function of $q$ in the region $q\in [q_{c5},
0.5]$. See the exemplified plots for $\alpha^2=0.7$ and 0.9 in the
right panel of Fig. \ref{fig:1}.

We summarize the connections between the variation trend of the BDD
and the direction of information flow between the system and the
reservoir as follows: for $\alpha^2\in (0, 0.382]$, the BDD can
always be enhanced by the backflow of the previously lost
information, while for $\alpha^2\in(0.382,0.5]$ ($\alpha^2\in
(0.5,1)$), it is enhanced with the information leaking into the
reservoir in the region of $q\in[q_{c3}, 0.5]$ ($q\in
[q_{c5},0.5]$), and by the information backflow otherwise.

\section{Explicit examples}\label{sec:5}
In this section, we illustrate through two examples the main
findings of this paper.

\subsection{Lorentzian spectral density reservoirs}

The first example we considered is the Lorentzian reservoir with
spectral density of the following form \cite{master}
\begin{equation}\label{eq5-1}
 J(\omega)=\frac{1}{2\pi}\frac{\gamma_0 \lambda^2}
           {(\omega-\omega_0)^2+\lambda^2},
\end{equation}
where $\lambda$ denotes spectral width of the reservoir and is
related to the reservoir correlation time via $\tau_B\approx
\lambda^{-1}$, while $\gamma_0$ denotes decay rate of the qubit's
excited state in the Markovian limit of flat spectrum and is related
to the qubit relaxation time via $\tau_R\approx\gamma_0^{-1}$.

For this reservoir, the factor $q(t)$ is given by \cite{Bellomo}
\begin{equation}\label{eq5-2}
 q(t)=e^{-\lambda t}\left(\cosh\frac{dt}{2}
      +\frac{\lambda}{d}\sinh\frac{dt}{2}\right)^2,
\end{equation}
with $d=(\lambda^2-2\gamma_0\lambda)^{1/2}$. Then from Eq.
\eqref{eq3-7} we obtain
\begin{equation}\label{eq5-3}
 \Gamma(t)=\frac{2\gamma_0 \lambda}{\lambda+d\coth\frac{dt}{2}}.
\end{equation}
It implies that there are temporary appearance of negative
$\Gamma(t)$ only when $\lambda<2\gamma_0$, for which the reservoir
is said to be non-Markovian, and the previously lost quantum
information may be fed back into the system again.

\begin{figure}
\centering
\resizebox{0.43\textwidth}{!}{%
\includegraphics{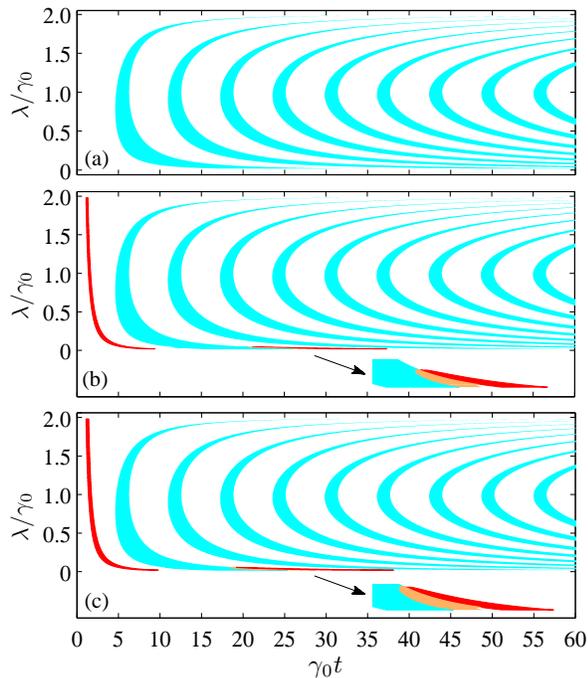}}
\caption{(Color online) Comparison between the direction of
         information flow and enhancement of the GQDs in Lorentzian
         reservoirs. For (a), $(\lambda, t)$ in the
         cyan shaded regions correspond to negative $\Gamma(t)$ which
         is a signature of information backflow, and the TDD with
         $\alpha^2\in (0,1)$, HDD with $\alpha^2\in(0,0.5]$, and BDD
         with $\alpha^2\in(0,0.382]$ can always be enhanced. For (b)
         and (c), the cyan (red) shaded regions correspond to negative
         (positive) $\Gamma(t)$ for which the HDD (b) and BDD (c) with
         $\alpha^2=0.7$ are enhanced, while the orange shaded regions
         correspond to negative $\Gamma(t)$ but  the HDD and BDD cannot
         be enhanced.} \label{fig:2}
\end{figure}

By choosing $\lambda/\gamma_0\in[0.02,1.98]$, we plotted in Fig.
\ref{fig:2} the parameter regions in which the GQDs can be enhanced.
For the TDD with $\alpha^2\in (0,1)$, HDD with $\alpha^2\in(0,0.5]$,
and BDD with $\alpha^2\in(0,0.382]$, they can always be enhanced by
the backflow information from the reservoir to the system. For the
HDD with $\alpha^2> 0.5$ and BDD with $\alpha^2>0.382$, one can see
that although there are $(\lambda,t)$ regions (the orange shaded
areas) in which they are degraded by the backflow information, and
regions (the red shaded areas) in which they are enhanced with the
information losing into the reservoir, they are in fact very narrow.
In most of the $(\lambda,t)$ regions, the temporary flow of
information from the reservoir back to the system can enhance the
values of them.

\subsection{Ohmic-like spectral density reservoirs}
The second type of structured reservoir we considered has the
Ohmic-like spectral density of the form \cite{Leggett}
\begin{equation}\label{eq5-4}
 J(\omega)=\eta\omega^s\omega_c^{1-s} e^{-\omega/\omega_c},
\end{equation}
where $\omega_c$ is the cutoff frequency, and $\eta$ the
dimensionless coupling constant. Their inverse are related to the
reservoir correlation time $\tau_B$ and the qubit relaxation time
$\tau_R$ via $\tau_B\simeq\omega_c^{-1}$ and $\tau_R\simeq
\eta^{-1}$. This reservoir is also said to be sub-Ohmic for $0<s<1$,
Ohmic for $s=1$, and super-Ohmic for $s>1$.

\begin{figure}
\centering
\resizebox{0.43\textwidth}{!}{%
\includegraphics{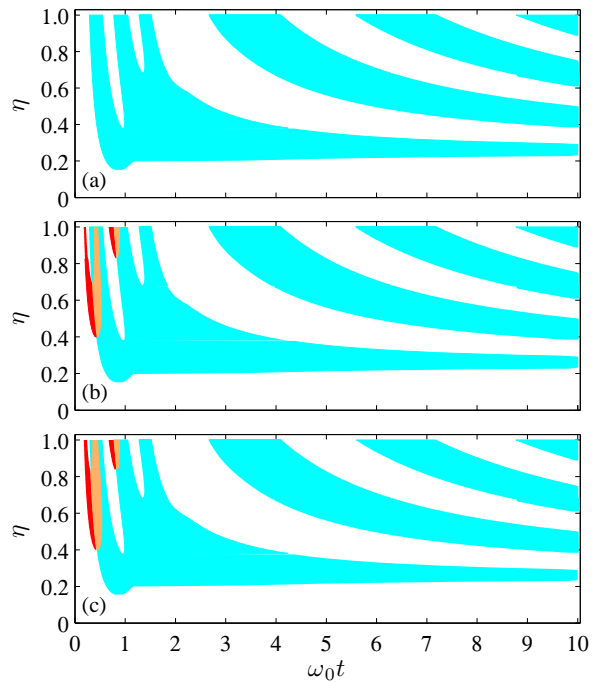}}
\caption{(Color online) Comparison between the direction of
         information flow and enhancement of the GQDs in super-Ohmic
         reservoirs with $s=3$ and $\omega_c= 2\omega_0$. For (a),
         $(\lambda, t)$ in the cyan shaded regions
         correspond to negative $\Gamma(t)$ which is a signature of
         information backflow, and the TDD with $\alpha^2\in (0,1)$,
         HDD with $\alpha^2\in(0,0.5]$, and BDD with $\alpha^2\in(0,0.382]$
         can always be enhanced. For (b) and (c), the cyan (red) shaded
         regions correspond to negative (positive) $\Gamma(t)$ for which
         the HDD (b) and BDD (c) with $\alpha^2=0.7$ are enhanced, while
         the orange shaded regions correspond to negative $\Gamma(t)$ but
         the HDD and BDD cannot be enhanced.} \label{fig:3}
\end{figure}

For the present case, there is no analytical solution for $q(t)$. In
the following, by fixing $s=3$ (i.e., we consider the super-Ohmic
reservoir) and $\omega_c= 2\omega_0$, we solved Eqs. \eqref{eq3-4}
and \eqref{eq3-7} numerically, and displayed the corresponding
results with $\eta \in [0.02,1]$ in Fig. \ref{fig:3}. From the plots
one can see that the information backflow always induces enhancement
of the TDD. For the HDD and BDD, however, there are very narrow
$(\lambda,t)$ regions (the orange shaded areas) in which they are
degraded by the information backflow, and very narrow regions (the
red shaded areas) in which they are enhanced by the information
loss. In a wide regime of $(\lambda,t)$, they are still be enhanced
by the information flowing from the reservoir back to the system.

It should be note that Fig. \ref{fig:3} is plotted with $\omega_0 t
\in [0, 10]$. When we extend it to a more wide region, we can find
there are also temporary enhancement of the HDD and BDD with the
information leaking into the reservoirs in the weak-coupling regime.
For other values of $\omega_c$, we also found similar connections
between the direction of information flow and enhancement of the
GQDs. For concise presentation, we did not plot the corresponding
figures here.

\section{Summary and discussion}\label{sec:6}
In summary, we have investigated evolutions of the GQDs for a pair
of qubits interacting independently with their own zero-temperature
bosonic structured reservoirs. The discord measures we adopted are
the well-accepted TDD, HDD, and BDD. By solving analytically their
dependence on a time-dependent factor $q(t)$ whose derivative
determines the non-Markovian character of the system dynamics, we
showed that the variation trend of the three GQDs are intimately
related to the direction of information flow between the system and
the reservoir. We identified explicitly the family of two-qubit states
for which the considered GQDs can be enhanced by the information
backflow from the reservoirs to the system, and states for which the
GQDs are enhanced with the information leaking into the reservoirs.

Moreover, by considering two explicit structured reservoirs with the
Lorentzian and Ohmic-like spectral density distributions, we showed
that although there are regions in which the information backflow
cannot enhance the magnitudes of HDD and BDD, and there are also
regions in which the HDD and BDD are enhanced by an increase in the
amount of information lost into the reservoirs, they are all very
narrow. In most of the parameter regions, they are still enhanced by
the information backflow. In this sense, non-Markovianity which
signifies a backflow of information from the environments to the
system, may be a potential resource deserved to be explored for
designing schemes by which the GQDs of open quantum systems can be
preserved or enhanced.

\section*{ACKNOWLEDGMENTS}
This work was supported by NSFC (11205121), and NSF of Shaanxi
Province (2014JM1008).

\newcommand{\PRL}{Phys. Rev. Lett. }
\newcommand{\RMP}{Rev. Mod. Phys. }
\newcommand{\PRA}{Phys. Rev. A }
\newcommand{\PRB}{Phys. Rev. B }
\newcommand{\PRE}{Phys. Rev. E }
\newcommand{\PRX}{Phys. Rev. X }
\newcommand{\NJP}{New J. Phys. }
\newcommand{\JPA}{J. Phys. A }
\newcommand{\JPB}{J. Phys. B }
\newcommand{\PLA}{Phys. Lett. A }
\newcommand{\NP}{Nat. Phys. }
\newcommand{\NC}{Nat. Commun. }
%

%

\end{document}